# MoS$_2$: a Choice Substrate for Accessing and Tuning the Electronic Properties of Graphene

*Chih-Pin Lu[1,2], Guohong Li[1], K. Watanabe[3], T. Taniguchi[3] and Eva Y. Andrei[1]*

[1] *Department of Physics and Astronomy, Rutgers University, Piscataway, New Jersey*

[2] *Department of Physics, National Taiwan University, Taipei 10617, Taiwan 08855, USA*

[3]*National Institute for Materials Science, 1-1 Namiki, Tsukuba, 305-0044, Japan*

One of the enduring challenges in graphene research and applications is the extreme sensitivity of its charge carriers to external perturbations, especially those introduced by the substrate. The best available substrates to date, graphite and hBN, still pose limitations: graphite being metallic does not allow gating, while both hBN and graphite having lattice structures closely matched to that of graphene, may cause significant band structure reconstruction. Here we show that the atomically smooth surface of exfoliated MoS$_2$ provides access to the intrinsic electronic structure of graphene without these drawbacks. Using scanning tunneling microscopy and Landau-level spectroscopy in a device configuration which allows tuning the carrier concentration, we find that graphene on MoS$_2$ is ultra-flat producing long mean free paths, while avoiding band structure reconstruction. Importantly, the screening of the MoS$_2$ substrate can be tuned by changing the position of the Fermi energy with relatively low gate voltages. We show that shifting the Fermi energy from the gap to the edge of the conduction band gives rise to enhanced screening and to a substantial increase in the mean-free-path and quasiparticle lifetime. MoS$_2$ substrates thus provide unique opportunities to access the intrinsic electronic properties of graphene and to study *in situ* the effects of screening on electron-electron interactions and transport.

The vulnerability of atomically thin layers such as graphene[1,2,3,4] to environmental disturbances has prompted an ongoing search for substrates that can support the material without perturbing its electronic structure. Graphite substrates were found to be by far the least invasive, making it possible to observe the intrinsic low energy spectrum of graphene by using scanning tunneling microscopy (STM), spectroscopy (STS)[4,5] and cyclotron resonance (CR)[6] measurements. However the metallic screening of graphite, which precludes control of the carrier-density by gating imposes severe limitations for both applications and fundamental studies. The alternative is to use insulating substrates, but the versatility gained comes at the price of enhanced sensitivity to surface corrugations,[7] and impurities,[8] which create electron-hole puddles that obscure the low energy electronic properties.[9] These perturbations can be mitigated by the use of atomically flat substrates such as mica[10] or hexagonal boron-nitride (hBN).[11,12] However a close match between the lattice structure of graphene and the substrate, as is the case of hBN or graphite, leads to a spatial modulation observed as a Moiré pattern in topography which can significantly perturb the electronic spectrum.[13,14,15] Here we show that atomically flat $MoS_2$ substrates provide access to the intrinsic band structure of graphene while at the same time they allow tuning via a gate voltage both the carrier density and the strength of screening.

$MoS_2$ is a semiconductor in the layered transition-metal-dichalcogenite family consisting of covalently bonded S-Mo-S sheets held together by the van der Waals force. Its weak interlayer coupling facilitates the extraction of ultra-thin layers by exfoliation. Bulk $MoS_2$ has an indirect band gap of 1.2 ~ 1.3 eV,[16] which due to quantum confinement, crosses over to a direct band gap of ~1.9 eV when the material is exfoliated down to a monolayer.[17] Thin layers of $MoS_2$ are well suited to serve as the channel material in field-effect transistor applications, exhibiting high mobility, almost ideal switching characteristics, and low standby power dissipation.[18,19, 20, 21,22] The fact that the position of $E_F$ can be promoted from the gap to the conduction band (CB) with modest gate voltages, allows tuning the screening properties of $MoS_2$ films from insulating to metallic. Further, the absence of dangling bonds and of surface states renders its surface inert, clean and minimally invasive.[23] The large lattice mismatch between $MoS_2$ and graphene, together with its chemical inertness and the tunable screening, renders $MoS_2$ ideally suited for gated STM/STS studies of graphene.[24]

We employed $MoS_2$ flakes exfoliated from bulk *2H*-$MoS_2$ crystals and deposited onto a 300 nm chlorinated $SiO_2$ substrate capping a degenerately p-doped Si gate. The thickness of the $MoS_2$ flakes, as measured by atomic force microscopy (AFM), ranged from monolayer to 40 nm. Exfoliated graphene was subsequently deposited on the $MoS_2$ flakes.[11] The devices were measured in a home-built STM.[25,26] Topography images were acquired in constant current mode. Differential conductance, dI/dV, which is proportional to the local density of states (LDOS),[27] was measured with a lock-in technique with fixed tip to sample distance. For details on sample fabrication and measurements see Supplemental Material S1.[28]

Figure1(a) illustrates the measurement setup and electrode configuration. The STM topography of graphene on $MoS_2$, Figure1(b), is compared to that on chlorinated $SiO_2$, Figure1 (c), and on two hBN substrates with and without moiré pattern, in Figure1 (e) right and left panels respectively. In Figure 1(d,f) we show the height histograms obtained from these topography images. The average surface corrugation, calculated from the standard deviation of Gaussian fits to the height histograms, is 27 ± 0.2 pm and 31 ± 0.1 pm on the $MoS_2$ on hBN substrates respectively. This is significantly smaller than the corrugation on $SiO_2$, ~234 ± 0.8 pm, in agreement with earlier reports.[7,29] We note that the presence of the moiré pattern ( Figure 1(e) right panel) leads to a substantially larger corrugation for graphene on hBN. In this case, in spite of the atomically flat hBN, the image exhibits a large periodic corrugation with an apparent height of ~ 0.4 nm[30] (Supplemental Material S2,[28]). The difference between the two images in Figure 1(e), both showing the topography of graphene on hBN, is due to the relative twist angle, $\phi$, between the lattice orientations of sample and substrate. The twist angle plays an important role at small lattice mismatch, $\delta = |a - a_S|/a_S$, as is the case for graphene on graphite,[13] $\delta \sim 0$, and on hBN,[29] $\delta \sim 1.8\%$. Here $a_S$ and $a$ are the lattice constants of the substrate and graphene respectively. At small $\delta$ a moiré superstructure forms with an angle dependent super-period,[29]

$$\lambda = a \frac{(1+\delta)}{\sqrt{2(1+\delta)(1-\cos\varphi) + \delta^2}}$$, which can open gaps[33,15] or introduce Van-Hove singularities[13] at energies corresponding to the superstructure reciprocal vector. Thus, depending on $\phi$, substrates with small $\delta$ can significantly disturb both the topography and the band structure even when they are atomically flat: the smaller $\phi$, the lower the energy at which band reconstruction sets in. In contrast, no reconstruction is expected for substrates with large $\delta$, as in the case of graphene on $MoS_2$ where $\delta \sim 0.3$.[31] As a result there is no need for special precautions about substrate orientation when depositing graphene on $MoS_2$.

Figure 2 shows the gate voltage ($V_g$) dependence of the STS spectra and of the Dirac point energy ($E_D$) measured with respect to $E_F$, which is taken as the zedro of energy. To understand the results for graphene on $MoS_2$ we first consider the case of graphene deposited on a chlorinated $SiO_2$ substrate, shown in Figure 2(b). In this case the data follow the typical square root dependence, $E_D = \hbar v_F \sqrt{\pi\alpha|V_g - V_0|}$, expected for the massless Dirac fermion spectrum of isolated graphene.[2,4] Here $\hbar$ is the reduced Planck constant and $v_F = 1.1 \pm 0.02 \times 10^6$ m/s is the Fermi velocity obtained, as discussed below, from the LL spectra. Fitting the data for these parameters we obtain the offset, $V_0 \sim 12$ V, and the reduced gate capacitance $\alpha \approx 7.3 \times 10^{10}$ cm$^{-2}$V$^{-1}$, from which we estimate the unintentional hole concentration, $n_0 = \alpha V_0 \sim 9 \times 10^{11}$ cm$^{-2}$. This value falls within the accepted range for graphene on SiO$_2$.[2] Expressing the reduced gate capacitance, $\alpha = \frac{\varepsilon \varepsilon_0}{de}$, in terms of the substrate thickness, $d = 300$ nm, and the dielectric constant, $\varepsilon$, we obtain $\varepsilon_{SiO2} \sim 4.1$ consistent with the accepted value for the dielectric constant of SiO$_2$. Here $e$ is the fundamental unit of charge and $\varepsilon_0$ the permitivity of free space.

Turning to graphene on MoS$_2$, Figure 2(c), we find that for $V_g < -10$ V the data are consistent with the expected square root dependence (solid line) calculated for the parameters obtained from the LL spectra: $v_F = 1.21 \pm 0.02 \times 10^6$ m/s, $\alpha \approx 6.6 \times 10^{10}$ cm$^{-2}$V$^{-1}$, and $V_0 \sim -4.5$ V. This gives an unintentional initial electron concentration of $n_0 = 7 \times 10^{11}$ cm$^{-2}$ and, using the thickness of the MoS$_2$ layer $d \sim 30$ nm, we obtain $\varepsilon_{MoS2} \sim 3.7$ for the dielectric constant of MoS$_2$. In contrast, for $V_g > -10$ V the gate dependence is significantly weaker, indicating that most of the gate induced charge is taken up by the MoS$_2$ substrate. Indeed from the finite field data, presented below, we find that in this regime only ~25% of the gate induced charge goes to the graphene layer, the rest being absorbed by the MoS$_2$ substrate.[21] This suggests that $E_F$ has entered the CB of MoS$_2$ at which point the gate induced shift in the position of $E_D$ is determined by the LDOS of the combined graphene/MoS$_2$ system, as illustrated in the inset of Figure 2(c). Naturally, as the LDOS in the CB of MoS$_2$ is larger than that of graphene most of the charge is absorbed by the former.

In the presence of a magnetic field normal to the graphene layer the spectrum breaks up into a sequence of LLs. Their evolution with field for the MoS$_2$ and SiO$_2$ substrates is shown in Figures 3(a) and 3(b) respectively. In both cases the sequence follows the field and level index (N) dependence characteristic of massless Dirac fermions:[4,32]

$$E_N = E_D \pm v_F\sqrt{2e\hbar|N|B} \quad N = 0, \pm 1, \pm 2 \dots \quad (1)$$

Here N > 0 (N < 0) and + (−) correspond to electron (hole) levels. Fitting the measured sequence to this expression we obtain $v_F(SiO_2)$ = 1.1 ± 0.02 × 10$^6$ m/s on chlorinated SiO$_2$ consistent with reported values.[2] In the case of graphene on MoS$_2$ where no prior measurements have been reported we find $v_F(MoS_2)$ = 1.21 ± 0.02 × 10$^6$ m/s. This gives the ratio $v_F(MoS_2)$ / $v_F(SiO_2)$ ~ 1.1 which is comparable to the ratio of the dielectric constants, $\varepsilon_{SiO2}/\varepsilon_{MoS2}$ ~ 1.1, consistent with the expected inverse dependence of $v_F$ on the dielectric constant.[33]

One of the prerequisites for observing well developed LLs is for the random potential to be smooth on the length scale of the cyclotron orbit, $l_c(B) = \sqrt{\frac{\hbar}{eB}} = \frac{25.64\text{nm}}{\sqrt{B}}$. The field at which LLs become resolved signals that the cyclotron orbit is sufficiently small to "fit" within the characteristic puddle size of a particular sample[34] and provides a direct measure of the quasiparticle mean free path (mfp) and sample quality. For graphene on MoS$_2$, Figure 3(a), the LLs are already resolved at 2 T indicating a characteristic puddle size exceeding $l_c(2\text{ T})$ ~ 18 nm. In contrast for the SiO$_2$ substrate, Figure 3(a), LLs only become distinguishable at 6 T, indicating smaller puddles, bounded by $l_c(6\text{ T})$ ~ 10 nm, and hence a shorter mfp. To test this conclusion we directly imaged the electron-hole puddles by mapping the spatial dependence of the doping level. For the SiO$_2$ substrate, Figures 3(c), the average puddle size obtained from the map, ~ 10 nm, is consistent with that obtained from the LL onset field. For the MoS$_2$ substrate, Figures 3(d), the larger puddle size obtained from the maps, ~ 30 nm, suggests that the LL would remain well resolved down to magnetic fields of 0.5 T, which are below the measurement range reported here. This value is slightly larger than for graphene on hBN[35], providing yet another indication of the high quality of the MoS$_2$ substrates.

To study the effect of doping on the LLs we measured the gate dependence of the spectra at fixed field, Figure 4(a) (10 T data shown in Supplemental Material S4,[28]). The pronounced staircase features reflect the high degeneracy of the LLs each of which can accommodate a carrier density of $D = 4\frac{B}{\phi_0} = 10^{11}$ B[T] states/cm$^2$, where $\phi_0 = 4.14 \times 10^{-11}$ Tcm$^2$, is the fundamental unit of flux.[4] In the process of gating, each LL as it is being filled pins the Fermi energy, and this produces the plateau features.[36] Since the width of a plateau, $\Delta V_g = D / \alpha$, reflects the gate voltage required to fill one LL, it provides a direct measure of α. In the regime where the MoS$_2$ substrate is insulating, $V_g < -10$ V, we find $\Delta V_g = 11.8$ V, which gives $\alpha \approx 6.6 \times 10^{10}$ cm$^{-2}$V$^{-1}$. For $V_g > -10$V, the plateaus become much wider, $\Delta V_g \sim 50$ V and $\alpha \approx 1.5 \times 10^{10}$ cm$^{-2}$V$^{-1}$, indicating the entry of $E_F$ into the of CB of MoS$_2$. At this point ~ 75% of the carriers introduced by the gate are taken up by the MoS$_2$ substrate which can now provide better screening of the random potential.

In Figure 4(b) we present the LL spectra for several values of $V_g$. Extracting $v_F$, Figure 4(c), we find that it is independent of $V_g$. Since $v_F$, is inversely proportional to $\varepsilon$ this implies that, in spite of the increased screening accompanying the entry of $E_F$ into the CB, the dielectric constant of MoS$_2$ is independent of doping for the range of $V_g$ employed here, consistent with recent theoretical work.[37]

To illustrate the effect of screening on the quasiparticle lifetime we compare in Figure 4(d) the linewidth, $\Delta E$, of the N = 0 LL in the unscreened ($V_g = -30$ V) and screened regimes ($V_g = +25$ V). Using a Gaussian fit we find ΔE ~ 53.5 mV for the unscreened case, which corresponds to a lifetime of $\tau \approx \hbar / \Delta E = 12$ fs and to a mfp of $l_{mfp} \sim v_F \tau \sim 15$ nm. In the screened regime we find a much narrower linewidth, $\Delta E \sim$ 28.2 mV, indicating significant reduction in scattering with correspondingly longer lifetimes ~ 23 fs and a mfp of $l_{mfp} \sim$ 28 nm. Interestingly the mfp obtained from the N

= 0 LL linewidth is comparable to the average puddle size, in Figure 3(c), indicating that for these samples the electron-hole puddles are the main source of scattering. A similar analysis of the N = 0 LL on chlorinated $SiO_2$ (Figure 3b) gives $\tau \sim$ 10 fs and $l_{mfp} \sim$ 11 nm comparable to the puddle sizes in this system, so that here also electron-hole puddles are the main source of scattering.

In summary, the quality of $MoS_2$ substrates as measured by the mfp is remarkably good: in the unscreened regime it is comparable to that in the best insulating substrates, hBN and chlorinated $SiO_2$, while in the screened regime it is larger still. The results presented here demonstrate that $MoS_2$ substrates are well suited for accessing the low energy electronic properties of graphene while at the same time providing great flexibility through controllable carrier densities and tunable screening.

Funding was provided by DOE-FG02-99ER45742 (E.Y.A, G.L), NSF DMR 1207108 (C.P.L) and National Science Council of Taiwan (C.P.L). We thank Ivan Skachko and Jinhai Mao for useful discussions.

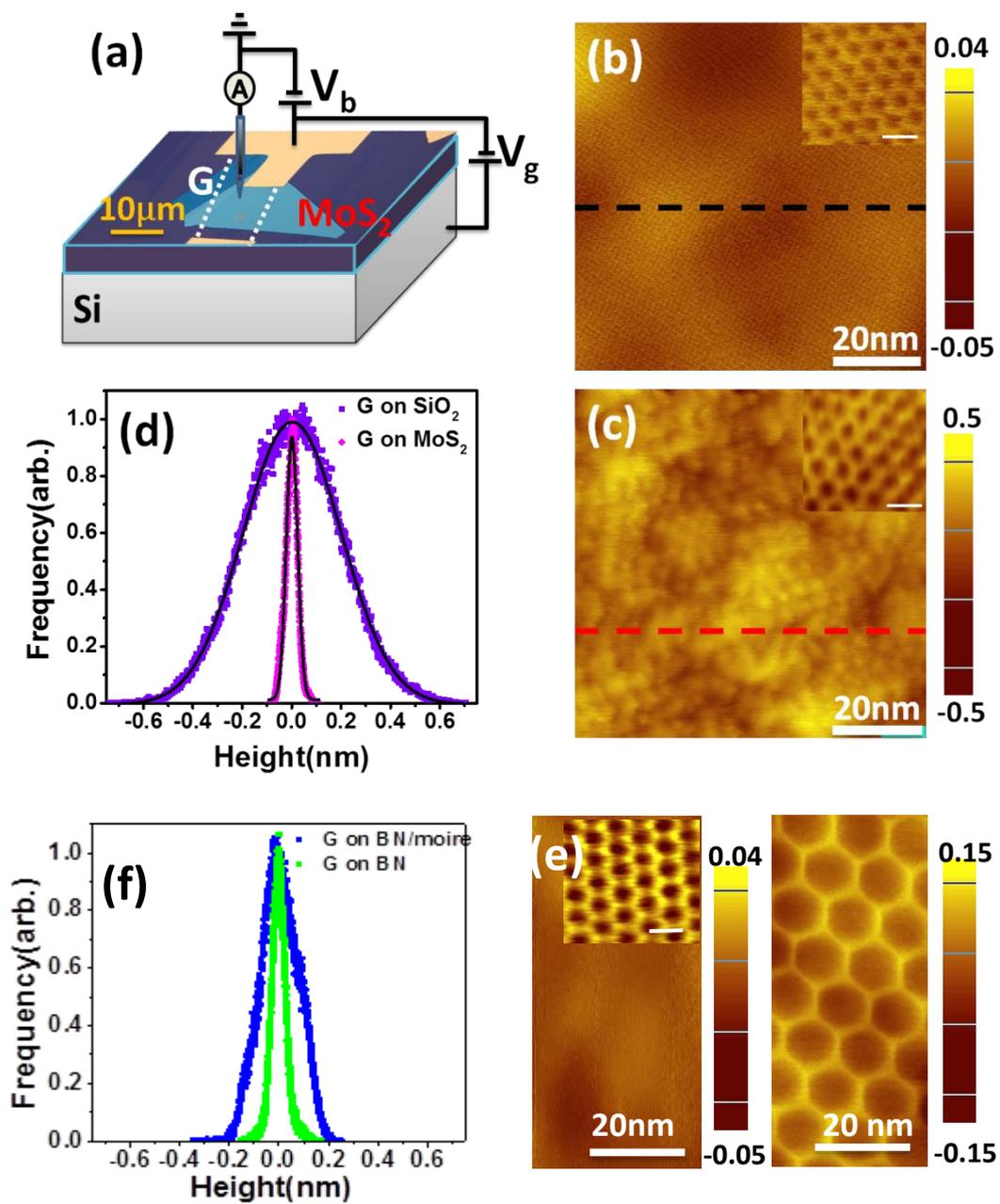

**Figure 1.** (a) Schematic of STS measurement setup showing the graphene sample (G) and MoS$_2$ substrate. The sample bias $V_b$ is applied between the STM tip and the sample. The edge of the graphene flake is marked by dashed lines. The back gate voltage $V_g$ is applied between the Si substrate and the top electrode. (b), (c) STM topography of graphene on MoS$_2$ and on chlorinated SiO$_2$ respectively. (d) Height histograms of the topography images in b and c. (e) Same as (b) for graphene on hBN with (right) and without (left) moiré pattern. (f) Height histograms of the topography images in (e). Image area 80 nm × 80 nm (b,c) and 40 nm × 80 nm(e). Insets in (b,c,e) represent zoom-in images with scale bar 0.3 nm. STS parameters: set point current $I = 20$ pA at $V_b = 0.4$ V. Height profiles shown in Supplemental material SM2 are taken along the dashed lines in (b),(c).

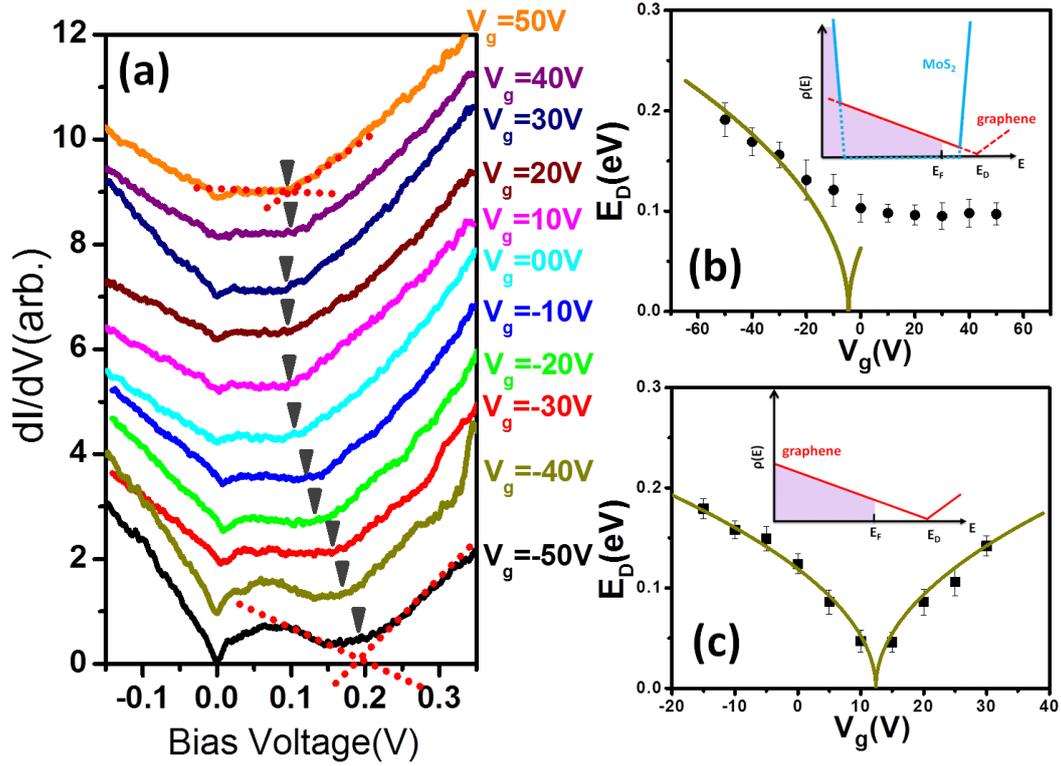

**Figure 2.** (a) Gate-voltage dependence of dI/dV spectra for graphene MoS$_2$. Arrows indicate the position of the Dirac point, $E_D$. Curves are vertically displaced for clarity. (b) Gate-voltage dependence of $E_D$ for graphene on a thin (~30 nm) MoS$_2$ substrate. The solid line represents a fit to the data as described in the text. Inset: Sketch of the combined DOS (solid lines) of graphene (dashed red lines) and MoS$_2$ substrate (dashed teal lines). (c) Same as (b) for graphene on a chlorinated SiO$_2$ substrate. Inset: Sketch of the DOS of p-doped graphene on an insulating substrate for energies within the substrate gap. STS parameters: set point current $I = 20$ pA at $V_b = 0.35$ V, modulation amplitude 5mV$_{rms}$.

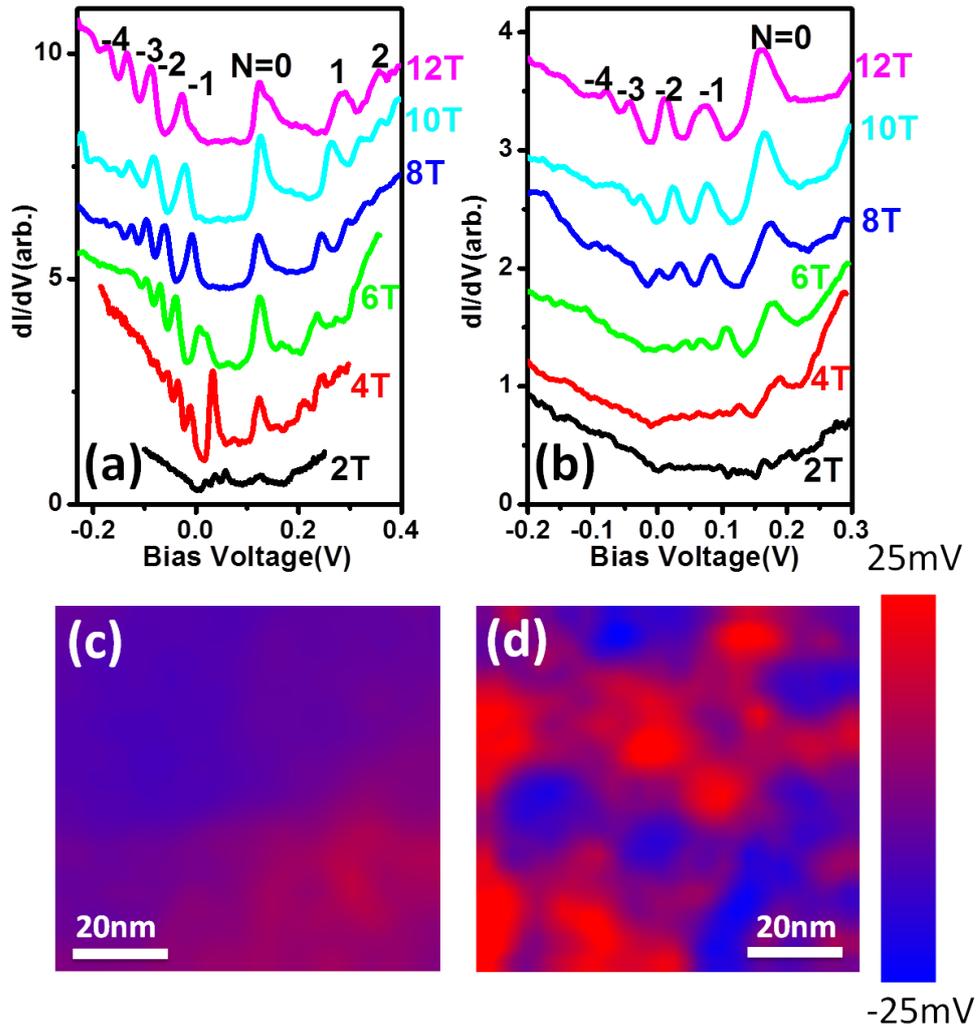

**Figure 3.** Field dependence of LL spectra. (a),(b) Spectra for graphene on MoS$_2$ and on SiO$_2$, respectively at $V_g = 10$ V. The LL indices $N = 0, \pm 1, \pm 2$..... are marked. (c),(d) Spatial variation of the $N = -1$ peak position at B = 10 T obtained from a *dI/dV* map at $V_g = 0$ V representing the doping inhomogeneity in graphene: blue (red) corresponds to hole (electron) doping. STS parameters: set-point $I = 20$ pA, $V_b = 0.35$ V, modulation amplitude 2mV$_{rms}$

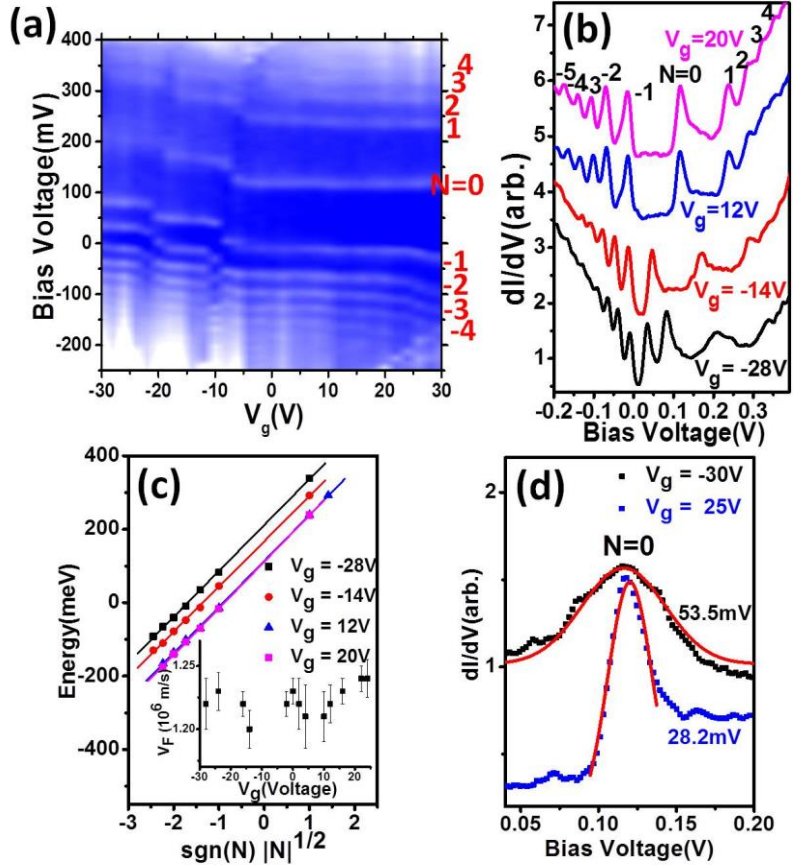

**Figure 4.** (a) Intensity map representing the gate dependence of the *dI/dV* spectra of graphene on $MoS_2$ at 8 T. Each vertical line corresponds to a LL spectrum at a particular gate-voltage. White staircase pattern corresponds to the LL peaks as indicated by the level-index N. STS parameters: set point current $I$ = 20 pA, $V_b$ = 0.4 V, ac modulation 5 mV$_{rms}$. (b) Gate-voltage dependence of LL spectra. (c) Gate-voltage dependence of LL peak sequences. Solid symbols are data points and lines are fits to equation 1. Inset: Gate dependence of Fermi velocity. (d) Comparison of the linewidth for the N = 0 LL in the unscreened ($V_g$ = −30 V) and screened ($V_g$ = 25 V) regimes. Symbols and lines represent data points and Gaussian fits respectively. Curves are offset for clarity.